\documentclass{elsart}
\usepackage[english]{babel}
\usepackage{times}
\usepackage{epsfig}
\usepackage{amsmath,amssymb}

\def\eq#1{\begin{equation}#1\end{equation}}
\def\al#1{\begin{align}#1\end{align}}
\def\singlefig{66mm}
\def\doublefig{66mm}

\date{12th March 2002}

\begin{document}

%\BackgroundEPS{fig/draft.eps}{25}{70}{2.5}

\begin{frontmatter}
\title{Scanning Lidar Based Atmospheric Monitoring for Fluorescence
Detectors of Cosmic Showers}
\author[PNG,IJS]{A.\ Filip\v ci\v c},
\author[PNG]{M.\ Horvat},
\author[PNG,IJS]{D.\ Veberi\v c\corauthref{cor}},
\corauth[cor]{Corresponding author.}
\ead{darko.veberic@ijs.si}
\author[PNG,IJS]{D.\ Zavrtanik},
\author[PNG,IJS]{M.\ Zavrtanik}
\address[PNG]{Nova Gorica Polytechnic, Vipavska 13, POB 301, SI-5001
Nova Gorica, Slovenia}
\address[IJS]{Jo\v zef Stefan Institute, Jamova 39, POB 3000, SI-1001
Ljubljana, Slovenia}

\begin{abstract}
Measurements of the cosmic-ray air-shower fluorescence at extreme
energies require precise knowledge of atmospheric conditions. The
absolute calibration of the cosmic-ray energy depends on the
absorption of fluorescence light between its origin and point of its
detection. To reconstruct basic atmospheric parameters we review a
novel analysis method based on two- and multi-angle measurements
performed by the scanning backscatter lidar system. Applied inversion
methods, optical depth, absorption and backscatter coefficient, as
well as other parameters that enter the lidar equation are discussed
in connection to the attenuation of the light traveling from shower to
fluorescence detector.
\end{abstract}

\end{frontmatter}

\section{Introduction}

Contemporary fluorescence experiments (Fly's Eye \cite{balt}, HiRes
\cite{abu}, P.\ Auger \cite{zavr}) studying cosmic rays with energies
near $10^{20}\,e$V detect fluorescence light produced along the
air-shower volume. As a cosmic-ray induced particle cascade develops
in the atmosphere it dissipates much of its energy by exciting and
ionizing air molecules. The cascade particles along the shower axis
are limited to a narrow lateral distribution where excited nitrogen
molecules are fluorescing in the near-UV spectral band. The efficiency
of the process is rather small. However, the vast number of emitting
particles in high energy shower makes this source of radiation highly
significant. Ultimately, this electromagnetic (EM) cascade dissipates
much of the primary particle's energy. Fluorescence light is emitted
isotropically with an intensity proportional to the number of charged
particles in the shower.
%
%Because fluorescence detection is essentially a calorimetric technique
%it is primarily sensitive to the electromagnetic (EM) component of the
%shower.
%
EM component and hence the total number of low energy EM particles is
in turn fairly accurately proportional to the energy of the primary
particle. Thus, the calorimetric measure of the total EM shower
energy \cite{design} is proportional to the integral of EM particle
density $N_{\text{em}}$ along the shower direction $x$,
\eq{
E_{\text{em}}=K\int N_{\text{em}}(x)\,\d x
\label{eem}
}
with $K\approx2.2\,$M$e$V\,cm$^2$/g, where $x$ is measured in units of
longitudinal air density (g/cm$^2$). $E_{\text{em}}$ is a lower bound
for the energy of the primary cosmic ray. The lower portion of shower
development is usually obscured by the ground so that EM cascade
reaching below ground is included by fitting a functional form to the
observed longitudinal profile and integrating the function past
surface depth. The number of photons $N_{\text{ph}}$ reaching
fluorescence detector (FD) is proportional to EM particle density
$N_{\text{em}}(x)$ at the point of production $x$, so that in turn
\eq{
N_{\text{em}}(x)\propto\frac{N_{\text{ph}}R^2(x)}{T(x)},
\label{particle_dens}
}
with $R(x)$ being distance between shower point $x$ and FD. Light
originating within the shower is certainly affected by the absorption
and scattering on molecules and aerosols in the atmosphere. The number
of detected photons is thus reduced due to non-ideal atmospheric
transmission $T(x)<1$, where
\eq{
T(x)=\exp\left[-\int_0^x\!\!\alpha(r)\,\d r\right]=\e^{-\tau(x)},
\label{transmission}
}
with $\alpha(r)$ volume extinction coefficient along the
line-of-sight, and $\tau(x)$ the resulting atmospheric optical depth
(OD) to the shower point $x$.

In this sense, the atmosphere can be treated as an elementary-particle
detector. However, weather conditions change the atmospheric
transmission properties dramatically resulting in strongly
time-dependent detection efficiency. Therefore, an absolute
calibration system for fluorescence light absorption is an essential
part of FD \cite{arcon,bird}.

Eq.~\eqref{transmission} is a basis for the fluorescence-detector
energy calibration. Apart from the Pierre Auger Observatory, all
existing fluorescence-based experiments have suffered from a lack of a
sufficient atmosphere monitoring system. Weather conditions in a
desert-like atmosphere were expected to be stable enough, so that the
standard attenuation-length profile should suffice to reconstruct the
total EM shower energy, Eq.~\eqref{eem}, with a controllable
precision. However, this has turned out not to be the case, especially
for rare events with energies above $10^{19}\,e$V. In addition, the
energy reconstruction is obscured by \v Cerenkov radiation in the
lower part of the air-shower which can not be separated from
fluorescence. For more than a half of the highest-energy events
measured so far the atmosphere properties are not known well enough to
accurately reconstruct the primary energy.
%
%Systematic shifts in energy could be as high as one order of
%magnitude.
%
To be able to provide adequate calibration, one has to measure
attenuation at the time of the event in the whole region of the
air-shower.

There is also a systematic discrepancy when comparing the cosmic-ray
spectra of fluorescence experiments with ground arrays. Their
compatibility can be established only with a shift in energy and flux
of one or the other. As in the case of fluorescence detectors, ground
arrays have their own problems with the energy determination and are
much more dependent on air-shower simulations. At present, it is not
known whether the discrepancy is due to fluorescence-detector or
ground-array method, or both. Therefore, it is of utmost importance to
have \textit{in situ} atmosphere monitoring system which is working
coherently with a fluorescence detector.

To lower primary cosmic ray energy uncertainties, the volume
extinction coefficient $\alpha(r)$ thus has to be well estimated over
almost whole detection volume of FD. In the case of the Pierre Auger
Observatory, the detection volume corresponds to ground area of
3000\,km$^2$ and height of $\sim15$\,km.

The paper is organized as follows. The first two sections are devoted
to introductory material on lidar measurement technique and a
description of our specific experimental setup. Then the atmospheric
model for simulation of lidar signals is presented and the signals are
evaluated by two well established inversion methods. Results of the
inversions are compared to the input model and conclusions on their
applicability are drawn. Next, improved approaches to FD calibration
based on scanning lidar system are proposed and evaluated on real
data obtained with our experimental setup.

\section{Lidar system}

One of the most suitable calibration setups for FD is the
backscattering lidar system, where a short laser light pulse is
transmitted from FD position in the direction of interest. With a
mirror and a photomultiplier tube, backscattered light is collected
and recorded as a function of time, i.e.\ as a function of backscatter
distance. Note that light from the lidar source traverses both
directions, so that in case of matching laser and fluorescence light
wavelength, OD for lidar light sums to twice the OD for
fluorescence. The lidar equation \cite{coll} describes the received
laser power $P(r)$ from distance $r$ as a function of volume
extinction coefficient $\alpha(r)$ and backscattering coefficient
$\beta(r)$,
\begin{equation}
P(r)=P_0\frac{ct_0}{2}\beta(r)\frac{A}{r^2}\,\e^{-2\tau(r)}.
\label{lidar-equation}
\end{equation}
$P_0$ is the transmitted laser power and $A$ is an effective receiving
area of the detector, proportional to the area of the mirror and
proportional to an overlap between its field of view with the laser
beam. $t_0$ is laser pulse duration. As seen from
Eq.~\eqref{particle_dens}, measurement precision of $\alpha$ and
corresponding $\tau$ directly influences the precision of primary
particle energy estimation.

Simple as it may look, the lidar equation~\eqref{lidar-equation} is
nevertheless difficult to solve for two unknown variables, $\alpha(r)$
and $\beta(r)$. All existing analysis algorithms (Klett \cite{klet},
Fernald \cite{fern}, and their respective variations) reviewed in one
of the following sections are based on an experimental setup with
static beam direction. This leads to ambiguity in determination of
$\alpha(r)$ and $\beta(r)$ which can not be resolved without
additional assumptions about atmospheric properties. At the FD
experimental sites the atmosphere can be assumed to be almost
horizontally invariant. In this case, there is an additional
constraint when comparing signals coming from different directions,
which solves the lidar equation for $\alpha(r)$ and $\beta(r)$
unambiguously. Even at this point, it can be stated that the need for
steerable (scanning) lidar setup is unavoidable for proper solution of
lidar equation.

\section{Experimental setup}

The lidar system used for verification of the analysis method is based
on the Continuum MiniLite-1 frequency tripled Nd:YaG laser, which is
able to transmit up to 15 shots per second, each with energy of 6\,mJ
and 4\,ns duration (1.2\,m). The emitted wavelength of 355\,nm is in
the $300-400$\,nm range of fluorescence spectrum. The receiver was
constructed using $80$\,cm diameter parabolic mirror with focal length
of 41\,cm. The mirror is made of aluminum coated pyrex and protected
with SiO$_2$.

\begin{figure}[t]
\centering
\epsfig{file=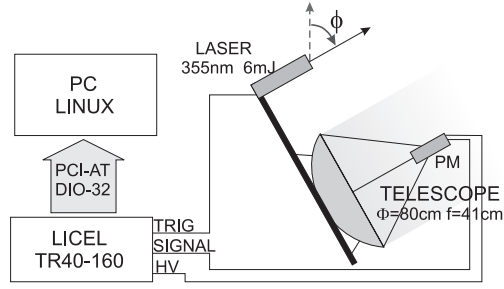,width=\singlefig,clip=}
\caption{Schematic view of the lidar system. A mirror of 80\,cm
diameter and a UV-laser head are mounted on the steerable
mechanism. The LICEL TR40-160 receives the trigger from the laser and
the signal from Hammamatsu R7400 phototube. The Linux-PC controls the
LICEL digitizer through PCI-DIO-32HS Digital Input/Output card. The
steering motors are controlled through RS-232 port. Zenith angle is
denoted by $\phi$.}
\label{f:lidar}
\end{figure}

The backscattered light is detected by a Hammamatsu R7400
photomultiplier with operating voltage up to 1000\,V and gain of
$10^5$ to $10^6$. To suppress background, a broadband UG-1 filter with
60\% trasmitance at 353\,nm and FWHM of 50\,nm is used. The distance
between laser beam and the mirror center is fixed to 1\,m, and the
system is fully steerable with $0.1^\circ$ angular resolution.

The signal is digitized using a three-channel LICEL transient recorder
TR40-160 with 12\,bit resolution at 40\,MHz sampling rate with 16k
trace length combined with 250\,MHz photon counting system. Maximum
detection distance of the hardware is thus, with this sampling rate
and trace length, set to 60\,km. However, in real measurements,
atmospheric features up to 30\,km only are observed. LICEL is operated
using a PC-Linux system through a National Instruments digital
input-output card (PCI-DIO-32HS) with Comedi dri\-vers \cite{comedi}
and a ROOT interface~\cite{root}.

\section{Lidar simulation with specific atmospheric model}
\label{s:simulation}

In a low opacity atmosphere the attenuation and the backscattering
coefficient can be written as a sum of contributions from two
independent components,
\begin{subequations}
\al{
\alpha(h)&=\alpha_{\text{m}}(h)+\alpha_{\text{a}}(h),
\\
\beta(h)&=P_{\text{m}}(180^\circ)\alpha_{\text{m}}(h)+
P_{\text{a}}(180^\circ)\alpha_{\text{a}}(h).
}
\label{alpha_beta}
\end{subequations}
where $\alpha_{\text{m}}$ and $\alpha_{\text{a}}$ correspond to
molecular and aerosol attenuation, respectively. The aerosol phase
function $P_{\text{a}}(180^\circ)$ for backscattering has, apart from
the wavelength, also a strong dependence on the optical and geometrical
properties of the aerosol particles. Nevertheless, at wavelength of
355\,nm, values in the range 0.025 and up to 0.05\,sr$^{-1}$ can be
assumed \cite{coll} for aerosol phase function
$P_{\text{a}}(180^\circ)$. The angular dependence of molecular phase
function is defined by the Rayleigh scattering theory, where
$P_{\text{m}}(180^\circ)=3/8\pi\,\text{sr}^{-1}$.

For simulation purposes, the elevation dependence of the extinction
coefficients is modelled as following,
\begin{subequations}
\al{
\alpha_{\text{m}}(h)&=\frac1{L_{\text{m}}}
\e^{-h/h_{\text{m}}^0},
\\
\alpha_{\text{a}}(h)&=\frac1{L_{\text{a}}}\begin{cases}
1, &h<h_{\text{x}}
\\
\e^{-(h-h_{\text{x}})/h_{\text{a}}^0},&h\geq
h_{\text{x}}.
\end{cases}
}
\label{ext-model}
\end{subequations}
where $L_{\text{m}}$ and $L_{\text{a}}$ are the molecular and aerosol
attenuation lengths at ground level, and $h_{\text{m}}^0$ and
$h_{\text{a}}^0$ are the molecular and aerosol scale height,
respectively. An additional mixing height $h_{\text{x}}$ is set up for
aerosols, assuming uniform concentration near the ground level and
continuous transition into exponential vanishing for
$h>h_{\text{x}}$. The following values of the parameters are used:
$L_{\text{m}}=15\,$km, $h_{\text{m}}^0=17.5\,$km,
$L_{\text{a}}=2\,$km, $h_{\text{x}}=0.8\,$km, and
$h_{\text{a}}^0=1.4\,$km.

The atmospheric model, Eq.~\eqref{ext-model}, serves as a testing
ground for two widely used reconstruction methods presented in the
next section. A comparison with reconstruction of the real atmosphere
yields insight into the common problems of the lidar field. The
Poissonian statistics of photon counting and multiplying, background
noise, and effects of digitalization have been taken into account in
the generation of the simulated lidar signals and under inspection
match those observed in the real lidar power returns.

\begin{figure}[t]
\centering
\epsfig{file=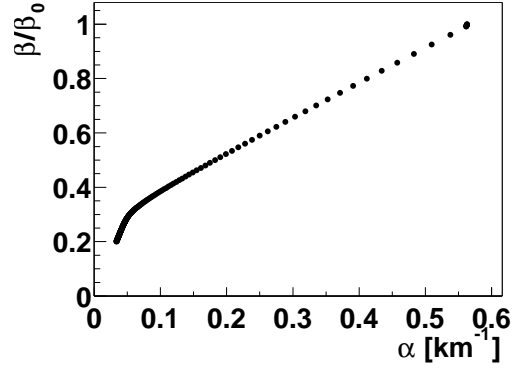,width=\singlefig,clip=}
\caption[]{Extinction--backscatter plot ($\alpha\beta$ diagram) for the
model atmosphere in Eq.~\eqref{ext-model}.}
\label{f:sim}
\end{figure}

Model in Eq.~\eqref{ext-model} is a valid approximation to the
atmospheric conditions found in real experiments. Although the
vertical variation of aerosol and molecular densities is quite simple,
the model still produces a non-trivial relation between total
attenuation $\alpha$ and total backscattering coefficient
$\beta$. Therefore, the dependence of $\beta$ as a function of
$\alpha$, shown in Fig.~\ref{f:sim}, cannot be well approximated by
some simple functional form.

\section{Reconstruction of a 1D atmosphere}
\label{s:1d}

Concentrating on a single shot lidar measurement, the optical
properties obviously have to be reconstructed in a 1D subspace of the
atmosphere. Rewriting the lidar equation \eqref{lidar-equation},
\eq{
P(r)=B\frac{\beta(r)}{r^2}\e^{-2\tau(r)}
\label{lidareq}
}
where the effective aperture of the system is gathered in the constant
$B$, an auxiliary $S$-function can be introduced,
\eq{
S(r)=\ln\frac{P(r)r^2}{P(r_0)r_0^2}=
\ln\left[\beta(r)/\beta_0\right]-2\tau(r;r_0).
\label{sfunct}
}
Note that $\tau(r;r_0)=\int_{r_0}^r\alpha(r')\,\d r'$ corresponds to
atmospheric OD between $r_0$ and $r$.

\subsection{Klett inversion}
\label{ss:klett}

Apart from the experimentally measured lidar power return $P(r)$, in
Eq.~\eqref{lidareq} there are two unknown quantities, $\beta$ and
$\alpha$ (or equivalently $\tau$), preventing the unique solution of
the lidar equation. Nevertheless, a simple, and sometimes physically
meaningful, assumption of proportionality between backscattering and
extinction,
\eq{
\beta(r)\propto\alpha^k(r),
\label{lidarrat}
}
allows for the transformation of the integral Eq.~\eqref{sfunct} to
the corresponding Ber\-noulli's differential equation with an existing
analytical solution. Direct application of the solution (forward
inversion) is numerically unstable, in some cases singular, and highly
sensitive to the signal noise \cite{klet,roca}. Klett's reformulation
\cite{klet} of the solution (backward inversion) avoids these
problems. The lidar backward inversion algorithm proceeds from the far
point of the measured signal $r_{\text{f}}$ to the near end,
\eq{
\alpha(r;\alpha_{\text{f}})=\frac{\e^{S(r)/k}}
{\e^{S_{\text{f}}/k}/\alpha_{\text{f}}+
\frac2k\int_r^{r_{\text{f}}}\e^{S(r')/k}\d r'},
\label{klett}
}
where $S_{\text{f}}=S(r_{\text{f}})$, and
$\alpha_{\text{f}}=\alpha(r_{\text{f}})$ is an estimate for the
attenuation at the far end of the data set. The reconstructed
attenuation $\alpha(r;\alpha_{\text{f}})$ is still a one-parameter
function of the unknown boundary attenuation value
$\alpha_{\text{f}}$, so that independent measurement, or suitable
approximation, is needed at the reference distance $r_{\text{f}}$. OD
can be expressed directly from Eq.~\eqref{klett},
\eq{
\tau(r;r_0,\alpha_{\text{f}})=\frac k2\ln\left[
\frac{k\e^{S_{\text{f}}/k}+2\alpha_{\text{f}}
\int_{r_0}^{r_{\text{f}}}\e^{S(r')/k}\d r'}
{k\e^{S_{\text{f}}/k}+2\alpha_{\text{f}}
\int_r^{r_{\text{f}}}\e^{S(r')/k}\d r'}
\right].
\label{klettod}
}

\begin{figure}[t]
\centering
\epsfig{file=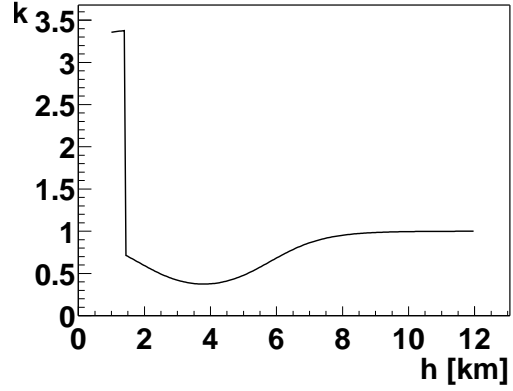,width=\singlefig,clip=}
\caption[]{Effective power $k$ in Eq.~\eqref{lidarrat} as obtained
from the model atmosphere in Eqs.~\eqref{alpha_beta} and
\eqref{ext-model}. Note that the discontinuity at
$h_{\text{x}}=1.4\,$km arises due to aerosol part of the model, up to
where aerosol concentration is kept constant.}
\label{f:klett_k}
\end{figure}

Klett's inversion method depends rather strongly on the assumed power
law proportionality in Eq.~\eqref{lidarrat}. In Fig.~\ref{f:klett_k},
a failure of this approximation is demonstrated for the specific
atmospheric model used for our simulations. The local value of the
exponent,
\eq{
k=\frac\alpha\beta\frac{\d\beta}{\d r}
\left[\frac{\d\alpha}{\d r}\right]^{-1},
\label{klett_k}
}
is shown to possess substantial range dependence. The main reason for
failure of the power law proportionality stems from the inequality of
the molecular and aerosol phase functions, $P_{\text{m}}(180^\circ)$
and $P_{\text{a}}(180^\circ)$, rendering the $\alpha$ and $\beta$
relationship dependent on the particular magnitude of both quantities,
and consequently range dependence (see Fig.~\ref{f:sim}). Therefore,
the best value of $k$ must be chosen using some \textit{ad hoc}
criterion.

\begin{figure}[t]
\centering
\epsfig{file=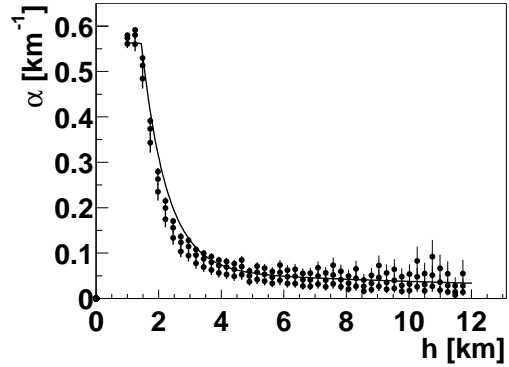,width=\singlefig,clip=}
\caption{Reconstructed attenuation $\alpha(h)$ from Klett's inversion
of the simulated vertical shot data as obtained by different boundary
values $\alpha_{\text{f}}$. Solutions with 0.5, 1, and 2 times the
correct $\alpha_{\text{f}}$ are plotted with dots. The actual model
$\alpha$ profile is drawn with solid line. Assuming range-independent
(constant) Klett's $k$, the best agreement between the reconstructed
and actual profile is achieved for $k\approx0.5$, therefore this value
is used for all tree plots.}
\label{f:klett}
\end{figure}

\begin{figure}[t]
\centering
\epsfig{file=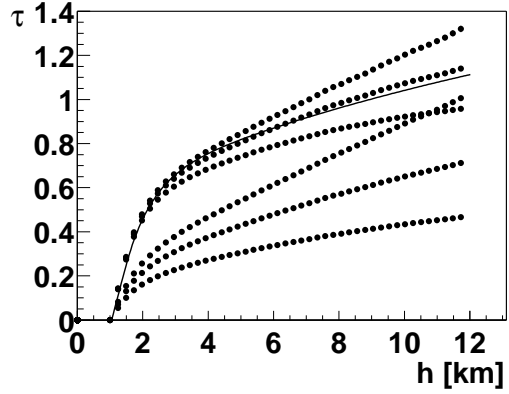,width=\singlefig}
\caption{Reconstructed optical depth $\tau(h;h_0)$ as obtained from
Klett's inversion in Fig.~\ref{f:klett} (with 0.5, 1, 2 times the
correct boundary $\alpha_{\text{f}}$). Upper three curves, with
different $\alpha_{\text{f}}$, are obtained using $k=0.5$ and lower
three with the most frequently used $k=1$. The actual $\tau$ profile
from our model is drawn with solid line and is seen to be well
approximated by the middle curve with $k=0.5$.}
\label{f:klett_od}
\end{figure}

Analyzing results in Fig.~\ref{f:klett}, presenting Klett inversion of
simulated lidar signals, it seems that the closest reconstruction of
the model profile is achieved with $k\approx0.5$. From
Fig.~\ref{f:klett_k}, showing local exponent $k$ obtained with use of
Eq.~\eqref{klett_k}, it can be seen that $k\approx0.5$ is observed
only in small interval around $4\,\text{km}$ whereas at other places
it is substantially larger. For $r>8\,\text{km}$, dominated by
molecular scattering it, slowly approaches the value of 1, most
commonly adopted in the literature. Nevertheless, as can be seen in
Fig.~\ref{f:klett_od}, reconstruction of OD with $k=1$ totally fails
to reproduce the correct answer. Surprisingly, in case of this
specific atmospheric model the most authentic result is obtained with
$k\approx0.5$.

Another drawback of the Klett's method is estimation of the extinction
$\alpha_{\text{f}}$ at the far end of the lidar return. In the case
that $r_{\text{f}}$ corresponds to a highly elevated point,
approximation
$\alpha_{\text{f}}\equiv\alpha_{\text{m}}(r_{\text{f}})$, i.e.\ the
extinction at that point is dominated by the molecular scattering,
yields quite reasonable results \cite{horv} with qualitative
convergence to the correct $\alpha$-profile. In general, for optically
dense atmosphere (e.g.\ presence of moderate haze) convergence of the
Klett's method is far more rapid as in clear, optically thin
case. However, sites for FD are usually chosen at locations with clear
and cloudless atmosphere. For horizontal lidar measurements (zenith
angle $\phi=90^\circ$) in a horizontally invariant atmosphere,
$\alpha_{\text{f}}$ can be estimated as the one that minimizes
extinction deviations from a constant value \cite{horv,yama}, i.e.
minimizes the functional
$\int_{r_0}^{r_{\text{f}}}[\alpha(r')-\alpha_{\text{f}}]^2\d r'$.

\subsection{Fernald inversion}
\label{ss:fernald}

Since the concentration of the molecules depends solely on the
thermodynamic parameters of the atmosphere, the Rayleigh scattering on
molecules is modeled separately on a basis of the meteorological data.
$\alpha_{\text{m}}(r)$ acquired in that way is inserted in
Eq.~\eqref{alpha_beta}. With an estimate for the molecular and aerosol
backscattering phase fraction,
$F=P_{\text{m}}(180^\circ)/P_{\text{a}}(180^\circ)$, and modified
$S$-function
\eq{
\tilde{S}(r)=S(r)+
2(F-1)\int_r^{r_{\text{f}}}\!\!\!\alpha_{\text{m}}(r')\,\d r',
\label{fernald}
}
the lidar equation can be solved for aerosol part
$\alpha_{\text{a}}(r)$ following the same steps as in Klett's version,
\eq{
\alpha_{\text{a}}(r)=-F\alpha_{\text{m}}(r)+\frac{\e^{\tilde S(r)}}
{\e^{\tilde S_{\text{f}}}/\tilde\alpha_{\text{f}}+
2\int_r^{r_{\text{f}}}\e^{\tilde S(r')}\d r'},
}
with
$\tilde\alpha_{\text{f}}=F\alpha_{\text{m}}(r_{\text{f}})+
\alpha_{\text{a}}(r_{\text{f}})$ and $\tilde S_{\text{f}}=\tilde
S(r_{\text{f}})=S(r_{\text{f}})$. In the same way OD is expressed as
\al{
\tau(r;r_0,\tilde\alpha_{\text{f}})=\,\,
&\frac12\ln\left[
\frac{\e^{\tilde S_{\text{f}}}+
2\tilde\alpha_{\text{f}}
\int_{r_0}^{r_{\text{m}}}\e^{\tilde S(r')}\d r'}
{\e^{\tilde S_{\text{f}}}+
2\tilde\alpha_{\text{f}}
\int_r^{r_{\text{m}}}\e^{\tilde S(r')}\d r'}
\right]+
\nonumber
\\
&+
(1-F)\int_{r_0}^r\!\!\alpha_{\text{m}}(r')\,\d r'.
}
Note that the Fernald procedure relies on three independently supplied
parameters: (i) an accurate estimate of molecular part of the
scattering $\alpha_{\text{m}}(r)$ along the whole range of interest,
(ii) total extinction at the far end $\tilde\alpha_{\text{f}}$, and
(iii) proper approximation for phase fraction $F$. As predicted by the
Mie theory, it is quite difficult to obtain reasonable values for the
latter. As for $\tilde\alpha_{\text{f}}$, conclusions are similar to
those of Klett's $\alpha_{\text{f}}$.

\begin{figure}
\centering
\epsfig{file=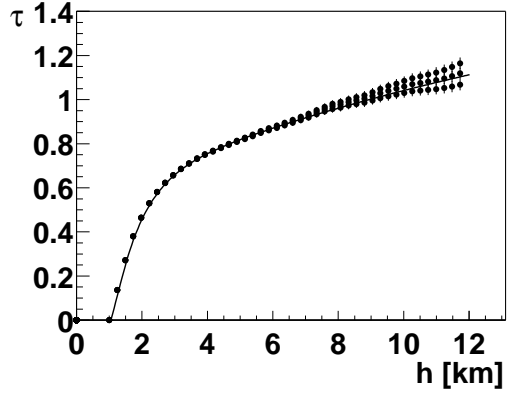,width=\singlefig,clip=}
\caption{Fernald inversion of simulated lidar signal. The correct
result is drawn in solid line. The three data sets are inversions with
$\alpha_{\text{a}}(r_{\text{f}})=0$ and
$\pm\alpha_{\text{m}}(r_{\text{f}})/2$. The phase fraction $F$ is kept
equal to the value used for generation of simulated lidar returns.}
\label{f:fernald}
\end{figure}

In Fig.~\ref{f:fernald} Fernald's inversion of simulated lidar return
is shown for different input values of
$\alpha_{\text{a}}(r_{\text{f}})$ that enter total extinction
$\tilde\alpha_{\text{f}}$. For upward pointing lidar measurements
vanishing aerosol concentration can be assumed at the far end of
atmosphere, i.e. $\alpha_{\text{a}}(r_{\text{f}})=0$. To test the
sensitivity of the reconstructed OD on this assumption, data sets with
$\alpha_{\text{a}}(r_{\text{f}})=\pm\alpha_{\text{m}}(r_{\text{f}})/2$
and therefore
$\tilde\alpha_{\text{f}}=(F\pm1/2)\alpha_{\text{m}}(r_{\text{f}})$,
are also plotted. $P_{\text{a}}(180^\circ)=0.025$\,sr$^{-1}$ is used
in the expression for phase fraction $F$. Comparing to the Klett's
method, which does not separate aerosol and molecular scattering, it
is not surprising that the variation of Fernald's results on boundary
parameters is somewhat weaker. Pinning the molecular part of
scattering undoubtedly stabilizes OD profiles obtained. Nevertheless,
Fernald's inversion still relies heavily on additional external
parameters that are usually difficult, if not impossible, to measure.

\section{Horizontally invariant atmosphere}
\label{s:h_inv}

Fluorescence detectors for cosmic showers are usually placed at
locations with specific atmospheric conditions. In case of the Pierre
Auger Observatory, the FD cameras are covering the lower part of the
atmosphere over an almost perfect $3000\,\text{km}^2$ plane
$1500\,\text{m}$ above the sea level with remarkable fraction of
cloudless days. Due to the high elevation and dry inland climate, an
optically thin atmosphere is expected. But, as noted before, in this
case convergence of Klett's method is slower and can lead to erroneous
estimates of OD. Based on that, and other peculiar problems of the
well established lidar inversion methods, a new approach with fewer
\textit{a priori} or hard-to-estimate input parameters is needed. Since
the lidar equation is not uniquely solvable, a minimal set of
assumptions needed for inversion has to be reconsidered. For a typical
FD site it is quite reasonable to assume weak horizontal variation of
the atmospheric optical properties. That is even more true for the
huge plane mentioned above, with hardly any changes in elevation and
vegetation coverage. Since the FD is exclusively operating at night,
only atmospheric conditions at that time have to be considered. The
mean night wind speeds do not exceed $12\,\text{km}/\text{h}$
\cite{baul}, so that particularly thin layer of aerosols close to the
ground is expected. At night, it is also expected that there will be a
low probability for formation of convective types of atmospheric
instabilities.

\subsection{Two-angle reconstruction}

Under the moderate assumptions presented above, optical parameters of
atmosphere that enter the lidar equation \eqref{lidareq} can be
assumed to possess only vertical variations, while being uniform and
invariant in the horizontal plane.

Thus, it makes sense to rewrite the range dependent $S$-function in
Eq.~\eqref{sfunct} in terms of height $h$ and geometric factor
$\xi=1/\cos\phi=\sec\phi$, when lidar shots with zenith angle $\phi$
are considered. The $S$-function becomes
\eq{
S(h,\xi)=\ln\left[\beta(h)/\beta_0\right]-
2\xi\,\tau(h;h_0)
\label{sfh}
}
with ``vertical'' OD $\tau(h;h_0)=\int_{h_0}^h\alpha(h')\d h'$ and
$\beta_0=\beta(h_0)$. After measuring two $S$-functions at different
zenith angles $\xi_1=1/\cos\phi_1$ and $\xi_2=1/\cos\phi_2$ and height
$h$, Eq.~\eqref{sfh} can be solved for the vertical OD,
\eq{
\tau(h)=-\frac12\frac{S(h,\xi_1)-S(h,\xi_2)}{\xi_1-\xi_2},
\label{od}
}
and the backscatter coefficient ratio,
\eq{
\frac{\beta(h)}{\beta_0}=\exp\left[
-\frac{\xi_2S(h,\xi_1)-\xi_1S(h,\xi_2)}{\xi_1-\xi_2}
\right].
\label{back}
}
Both quantities are directly proportional to the difference of two
$S$-functions at the same height and different angles. Therefore,
choosing a small separation between zenith angles, $\xi_1=\xi$ and
$\xi_2=\xi+\d\xi$, a differential form of Eq.~\eqref{od} can be
written,
\eq{
\tau(h)=-\frac12\frac{\partial S}{\partial\xi}\Big|_h.
\label{tau_diff}
}
Equivalently, the differential form of Eq.~\eqref{back} can be
obtained,
\eq{
\frac{\beta(h)}{\beta_0}=\exp\left[
S(h,\phi)-\xi\frac{\partial S}{\partial\xi}\Big|_h
\right].
\label{beta_diff}
}
Note that the OD is in that way determined up to the additive
constant, and the backscatter coefficient up to the multiplicative
factor. Nevertheless, both values should satisfy $S(h_0)=0$ and
$\tau(h_0)=0$.

Taking into account the Poissonian statistics of collected photons,
and neglecting all other sources of measurement uncertainties, a
relative error of the obtained OD at some height depends on the lidar
system parameters,
\eq{
\frac{\sigma_\tau}\tau=\frac{h/h_0}{2\tau\sqrt{\smash[b]{N_0\tilde\beta}}}\cdot
\frac1{|\xi_1-\xi_2|}\sqrt{\e^{2\xi_1\tau}+\e^{2\xi_2\tau}},
\label{err_tau}
}
as well as relative error of backscatter coefficient
\eq{
\frac{\sigma_\beta}\beta=\frac{h/h_0}{\sqrt{\smash[b]{N_0\tilde\beta}}}\cdot
\frac{\sqrt{\xi_2^2\e^{2\xi_1\tau}+\xi_1^2\e^{2\xi_2\tau}}}
{|\xi_1-\xi_2|},
\label{err_beta}
}
where $N_0$ is number of detected photons in the time interval
corresponding to the power return from height $h_0$, and
$\tilde\beta=\beta/\beta_0$.

\begin{figure}[t]
\centering
\epsfig{file=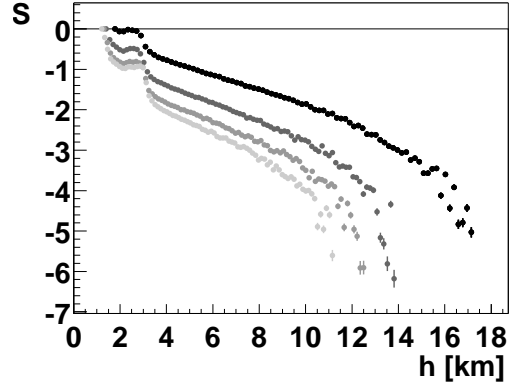,width=\singlefig,clip=}
\caption{$S$-function at few angles, $\phi=0^\circ$, i.e.\ $\xi=1$
(upper data set in black), $\phi=38^\circ$, $42^\circ$, and $47^\circ$
($\xi=1.27$, $1.35$, and $1.47$), in shades of gray (lower three data
sets).}
\label{f:s}
\end{figure}

\begin{figure}[t]
\centering
\epsfig{file=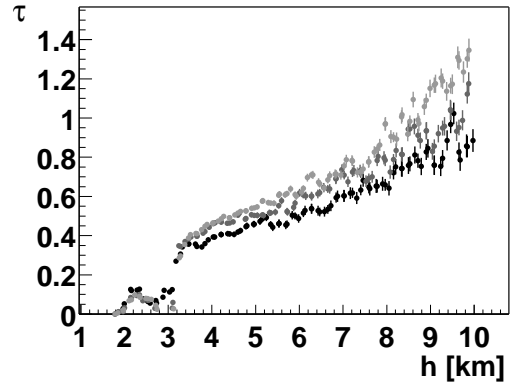,width=\singlefig,clip=}
\caption{Reconstructed optical depth (OD) $\tau$ from three pairs of
$S$-functions. In all pairs, $S_1$ corresponds to the $S$-function
with $\phi=0^\circ$ ($\xi=1$) and $S_2$ to the $S$-functions with
$\phi=38^\circ$, $42^\circ$, and $47^\circ$ ($\xi=1.27$, $1.35$, and
$1.47$), respectively from bottom to top.}
\label{f:od}
\end{figure}

\begin{figure}[t]
\centering
\epsfig{file=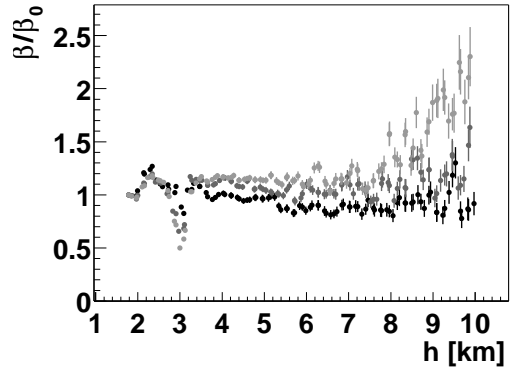,width=\doublefig,clip=}
\caption{Reconstructed backscatter coefficient $\beta(h)/\beta(h_0)$
from three pairs of $S$-functions. In all pairs, $S_1$ corresponds to
the $S$-function with $\phi=0^\circ$ ($\xi=1$) and $S_2$ to the
$S$-functions with $\phi=38^\circ$, $42^\circ$, and $47^\circ$
($\xi=1.27$, $1.35$, and $1.47$), respectively from bottom to top.}
\label{f:back}
\end{figure}

\begin{figure}[t]
\centering
\epsfig{file=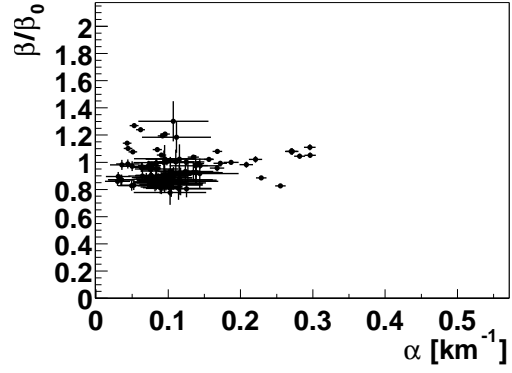,width=\doublefig,clip=}
\caption{$\alpha\beta$ diagram (extinction-to-backscatter plot) for
the pair where $S_1$ is taken at $\phi=0^\circ$ and $S_2$ at
$\phi=38^\circ$. Note that in Fig.~\ref{f:sim} the same diagram is
plotted also for model atmosphere.}
\label{f:ab}
\end{figure}

\begin{figure}[t]
\centering
\epsfig{file=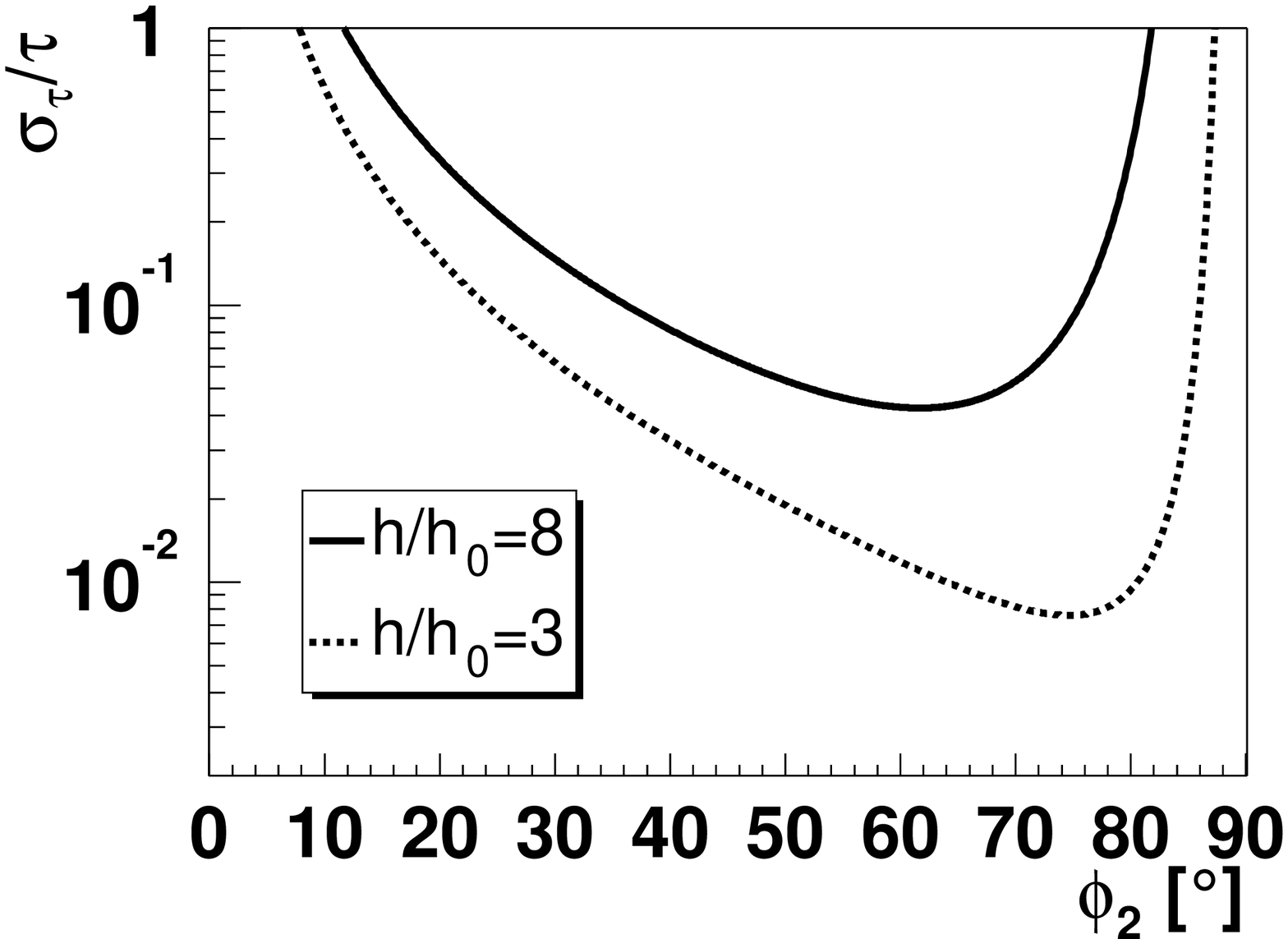,width=\singlefig,clip=}
\\[2mm]
\epsfig{file=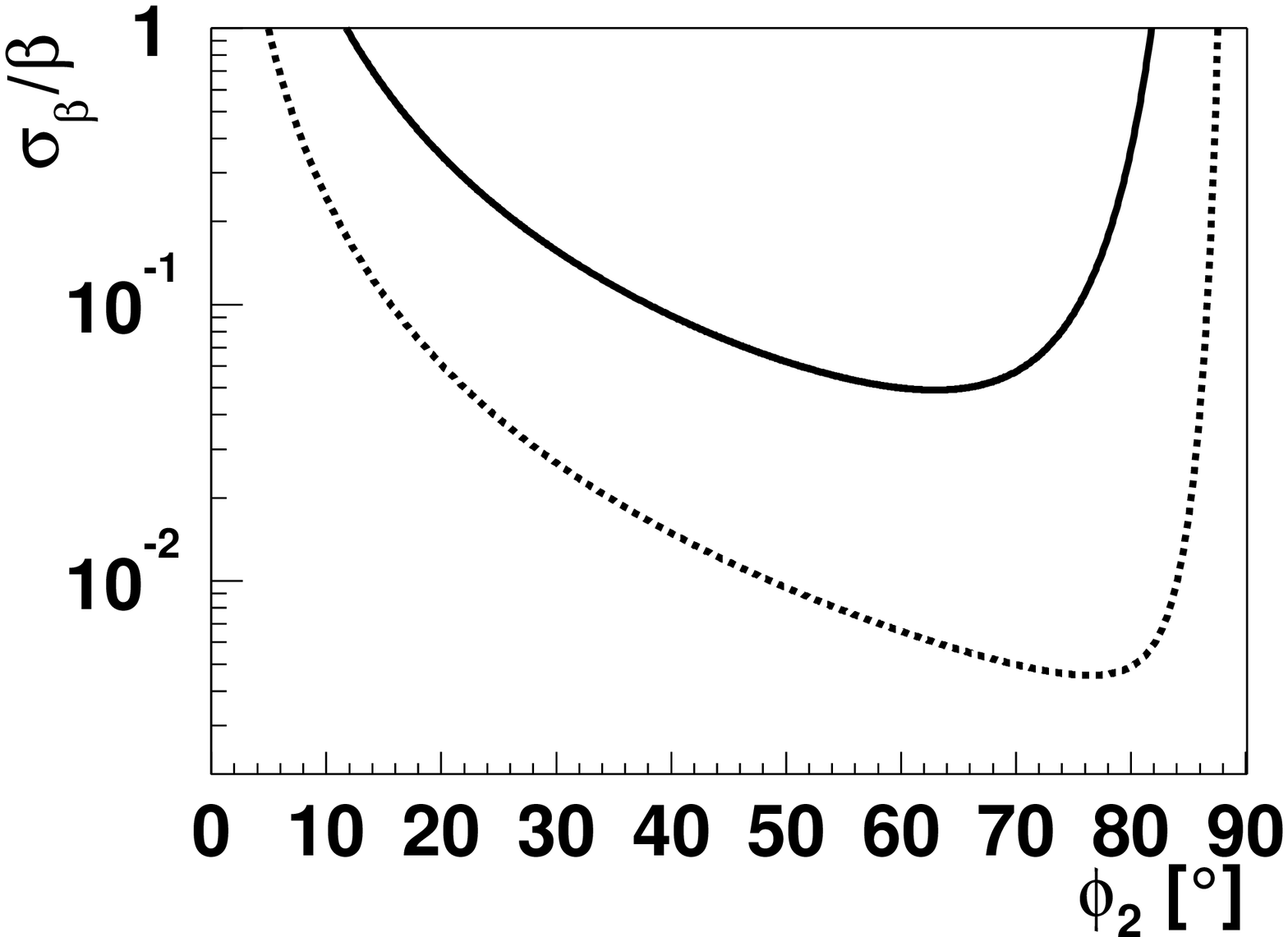,width=\singlefig,clip=}
\caption{Logarithmic plot of the relative deviation in OD,
$\sigma_\tau/\tau$ (upper panel), and backscattering coefficient,
$\sigma_\beta/\beta$ (lower pannel), vs.\ the second shot angle
$\phi_2$, in case that the first shot zenith angle is set to
$\phi_1=0^\circ$. Values $h/h_0=8$, $N_0=4\cdot10^6$, $\tau=1$, and
$\beta/\beta_0=0.6$ corresponding to the far point ($h\approx8\,$km)
in Fig.~\ref{f:od} have been assumed for parameters in
Eqs.~\eqref{err_tau} and \eqref{err_beta} (solid line). Values
$h/h_0=3$, $N_0=4\cdot10^6$, $\tau=0.4$, and $\beta/\beta_0=0.8$
corresponding to the near point ($h\approx2\,$km) are assumed for the
dashed curves. Note that $\phi=60^\circ$ corresponds to $\xi=2$.}
\label{f:err}
\end{figure}

In Fig.~\ref{f:s}, an example of $S$-functions and their zenith angle
variation is presented. All results are obtained from real lidar
measurements performed during few November nights in a typical urban
atmosphere (GPS location: 46$^\circ$04'35" N, 014$^\circ$29'05" E,
312\,m above sea level). For fixed primary azimuth angle
$\phi_1=0^\circ$ and three selected secondary angles
$\phi_2=38^\circ$, $42^\circ$, and $47^\circ$ results for OD
(Fig.~\ref{f:od}), backscatter coefficient (Fig.~\ref{f:back}), and
$\alpha\beta$ diagram (Fig.~\ref{f:ab}) are obtained from
corresponding $S$-functions in Fig.~\ref{f:s}. Due to presence of a
thin layer of optically thick haze at $h\approx3\,{\text{km}}$, a
drastic change in both OD and backscattering at that height is
observed. Since OD is well determined only up to an additive constant,
note that the variation of results for different $\phi_2$ is easily
produced by the inadequate determination of $S_0$, in other terms, by
variation of atmospheric optical properties at $h_0$. Compatible with
a scale height of $\sim18\,\text{km}$, the variation of backscattering
in Fig.~\ref{f:back} is slower as found in our model, generating a
gradual but still comparable $\alpha\beta$ diagram in Fig.~\ref{f:ab}.

In Fig.~\ref{f:err}, a logarithmic plot of the relative error in OD is
presented for typical lidar system parameters. First, the angle is
fixed to $\phi_1=0^\circ$ while the second one, $\phi_2$, is varied
from a vertical to an almost horizontal shot. It is hard to avoid the
fact that minimum error is produced with evaluation of two quite
considerably separated lidar shots, $\phi_2\approx70^\circ$. Even at
moderate elevations $h$ this can amount to large spatial separations
of the two points of lidar return, and thus the requirement of
horizontal invariance easily broken. In the case of an atmosphere,
that is slowly horizontally modulated, a more ``local'' approach to
the OD problem is needed.

\subsection{Multi-angle reconstruction}

For the ideal atmosphere, with true horizontal invariance, the $\xi$
dependence of $S$-function is particularly simple,
\eq{
S(h,\xi)=\ln[\beta(h)/\beta_0]-2\xi\,\tau(h;h_0),
\label{s_lin}
}
with the backscatter coefficient $\ln[\beta/\beta_0]$ as offset, and
OD $\tau$ as the slope of the resulting linear function in
$\xi$. Therefore, the optical properties of the atmosphere can be
alternatively obtained from the analysis of the $S$-function behavior
for scanning lidar measurements. Furthermore, disagreement of the
measured $S(\xi)$ profiles from the linear form is a suitable
criterion for detection of deviations from the assumed horizontal
invariance of the atmosphere.

A generalization of the two-angle equations \eqref{tau_diff} and
\eqref{beta_diff} to their differential counterparts strongly
suggested this way of reconstruction of optical properties, the
two-angle method being a mere two-point approximation of the linear
function in Eq.~\eqref{s_lin}. Taking into account quite substantial
uncertainties in $S(\xi)$ for single angle, the linear fit trough many
data points seems to yield superior results and the reconstruction is
no longer limited to two lidar shots, well-separated in angle. The
preferred horizontal invariance is not required to take place across
huge atmospheric volumes (as in case of $\phi_1=0^\circ$ and
$\phi_2=60^\circ$ shots), but has to be met only in relatively small
arc of interest where the continuous lidar scan is performed.

In the opposite case, when slow variation of atmospheric properties in
horizontal plane is allowed, Eq.~\eqref{tau_diff} is similar enough to
the renown 1D ``slope method'', based on assumption of small variation
of $\beta(r)$, or equivalently $\d\beta/\d r\approx0$. Bear in mind
that in method presented here the variation of $\beta$ with height can
be of any magnitude, as long as there are only modest variations in
the horizontal direction.

\begin{figure}[t]
\centering
\epsfig{file=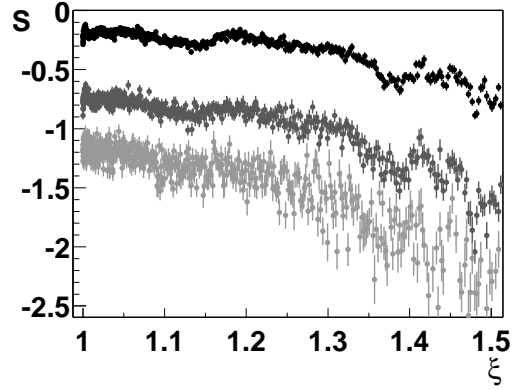,width=\doublefig,clip=}
\caption{Dependence of $S$-function on azimuth angle at various
heights $h=3.2$ (black), 5.6 (gray), and 8\,km (light gray), while
$h_0=3\,$km. Note that $\xi=1$ corresponds to $\phi=0^\circ$, and
$\xi=1.5$ to $\phi=48^\circ$.}
\label{f:s_xi}
\end{figure}

Opposite to Fig.~\ref{f:s}, in Fig.~\ref{f:s_xi} $S$-function profiles
with respect to zenith $\xi$ are drawn for fixed heights, starting
with $h=3.2\,$km and up to $7\,$km with $633\,$m step. Approximate
linear behavior is observed in few arc intervals, with narrow bands of
minute atmospheric shifts at $\xi=1.15$ and 1.38. Since these shifts
in profiles disappear when lifting $h_0$ from $3\,$km to $3.5\,$km,
they are obviously due to the distortions of atmosphere in the latter
interval, feature already observed in Fig.~\ref{f:od}.

\begin{figure}[t]
\centering
\epsfig{file=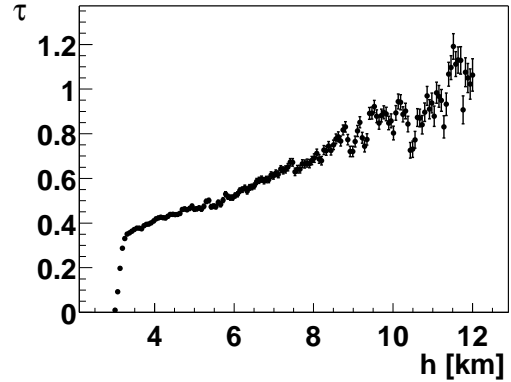,width=\doublefig,clip=}
\caption{Optical depth $\tau$ obtained by linear fits of angle dependence of
$S$-functions in Fig.~\ref{f:s_xi}.}
\label{f:multiod}
\end{figure}

In Fig.~\ref{f:multiod} the results of fitting and extraction of OD
are similar to the ones in Fig.~\ref{f:od}. Note that in both cases OD
is obtained relative to the $h_0$ point, so that the results may
differ up to some additive constant. Therefore, comparing both
figures, it is more accurate to concentrate on the same span of OD
within the $3.5\,$km to $9\,$km interval. Nevertheless, the range of
OD results with acceptable error bars is with multi-angle method
increased up to $12\,$km.

\begin{figure}[t]
\centering
\epsfig{file=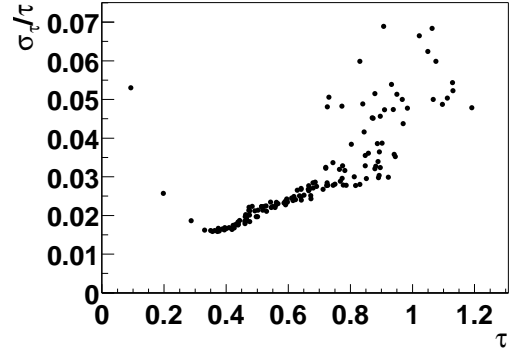,width=\doublefig,clip=}
\caption{Dependence of relative error in optical depth on depth
itself, $\sigma_\tau/\tau$. Data points and uncertainties are from
Fig.~\ref{f:multiod}.}
\label{f:err_tau}
\end{figure}

\begin{figure}[t]
\centering
\epsfig{file=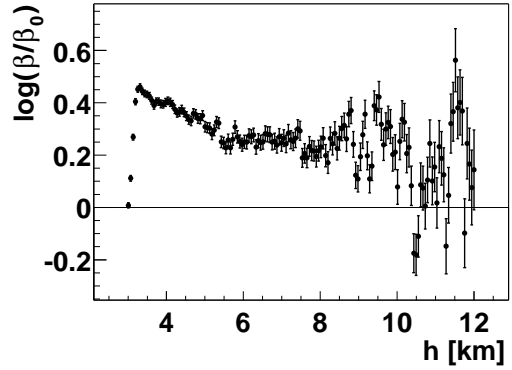,width=\doublefig,clip=}
\caption{Relative backscattering coefficient $\beta(h)/\beta_0$ from
$S$-functions in Fig.~\ref{f:s_xi}.}
\label{f:multi_beta}
\end{figure}

The relative error of OD in Fig.~\ref{f:err_tau} is needed for correct
estimation of shower energy uncertainty. It is kept below 6\% even for
the OD from the far points of the range, and below 3\% for modest
values of OD. Fig.~\ref{f:multi_beta}, with values for $\beta/\beta_0$
should be compared to Fig.~\ref{f:back}.

\section{Conclusions}

Inversion attempts of simulated lidar returns for atmosphere, modeled
by Eqs.~\eqref{ext-model}, show numerous drawbacks of established
numerical methods. For instance, Klett's and Fernald's method of
section \ref{ss:klett} and \ref{ss:fernald} do not satisfy the
specific requirements of FD calibration. While they may be useful for
qualitative reconstruction of atmospheric properties (spatial
haze/cloud distribution, cloud base etc.), they are not applicable for
absolute assessment of atmospheric transmission properties. There are
many reasons for this failure. One of them is certainly strong
dependence of obtained inversions on presumed extinction/backscatter
functional relation, Eq.~\eqref{lidarrat}, in case of Klett's method,
and assumed spatial dependence of Rayleigh scattering on molecules in
Fernald's case. Another issue is the extraordinarily difficult
measurement of far-side extinction rate $\alpha_{\text{f}}$, needed in
Eq.~\eqref{klett}, and phase fraction $F$, Eq.~\eqref{fernald}. We are
therefore forced to find better solutions, even at the expense of
adding scanning capabilities to an otherwise rigid lidar setup.

In contrast to that, based on the sole assumption of a horizontally
invariant (or at least horizontally slowly varying) atmosphere, the
two-, and especially the multi-angle, method presented in section
\ref{s:h_inv}, while simple in structure, nevertheless produce
reliable quantitative answers with small uncertainties (e.g., see
Figs.~\ref{f:od} and \ref{f:multiod}) to FD calibration questions. As
found by our investigation of first-run measurements, the relative
error of OD for distances up to 12\,km stay well below 6\%. This
number can be reduced even further by slow angular scanning and fast
multiple-shot averaging of lidar returns. Nevertheless, in that case
increased interaction between FD and lidar laser source, especially FD
blind time, has to be taken into account. Furthermore, concerning the
specific form of the atmospheric transmission entering
Eq.~\eqref{particle_dens}, they offer suitable starting ground for
development of methods that can considerably reduce systematic errors
of shower energy $E_{\text{em}}$ estimation with fluorescence
detectors.

In the case of strict horizontal invariance, both methods deliver
exact solutions of the lidar Eq.~\eqref{lidar-equation} with accuracy
of the results limited only by the quality of the measurement. In that
way, they offer reliable framework for study of the notorious
\textit{lidar ratio problem} (i.e., extinction to backscatter
codependency), widely discussed in the pure lidar community
\cite{earl}. Since, for example in the case of the Pierre Auger
Observatory, where optical properties have to be known over large
volumes of atmosphere, and a scanning lidar is therefore a necessity,
both mentioned methods represent natural first choice of data
analysis.

\begin{ack}
Authors would like to express gratitude to O.\ Ullaland for the
support and encouragement during our work. Authors also wish to thank
G.\ Navarra for assistance with EAS-TOP telescopes. This work has been
supported by the Slovenian Ministry of Education, Science, and Sport
under program No.~P0-0501-1540.
\end{ack}

\end{document}